\documentclass[aps, prd, reprint, twocolumn,nofootinbib]{revtex4}

\usepackage{color}
\usepackage{soul}
\usepackage{amssymb}
\usepackage{epsfig}
\usepackage{epstopdf}
\usepackage{graphicx,color}
\usepackage[usenames,dvipsnames]{xcolor}
\usepackage[section]{placeins}
\usepackage{amsmath}

\usepackage[normalem]{ulem}
\usepackage{booktabs}
\def\Journal#1#2#3#4{{#1} {\bf#2}, #3 (#4)}
\def\PRC{{\rm Phys. Rev.} C}

\def\la{\langle}
\def\ra{\rangle}

\def\be{\begin{equation}}
\def\ee{\end{equation}}
\def\bea{\begin{eqnarray}}
\def\eea{\end{eqnarray}}

\usepackage{xcolor}
\colorlet{darkgreen}{green!50!black}
\colorlet{orange}{red!50!yellow}
\colorlet{darkblue}{blue!60!black}
\colorlet{darkred}{red!80!black}

\usepackage{mathtools}
\usepackage{mathrsfs}
\usepackage{bm}
\usepackage{indentfirst}

\usepackage{xcolor}

\begin{document}
% \begin{frontmatter}
\title{Beam Spin Asymmetry in Electroproduction of  Pseudoscalar or Scalar Meson Production off the Scalar Target}
\author{Chueng-Ryong Ji$^a$}
\author{Ho-Meoyng Choi$^b$} 
\author{Andrew Lundeen$^a$}
\author{Bernard L. G. Bakker$^c$}
\address{$^a$Department of Physics, North Carolina State University, Raleigh, North Carolina, 27695-8202, USA}
\address{$^b$Department of Physics, Teachers College, Kyungpook National University, Daegu, Korea 41566}
%\address{$^c$Department of Physics and Astrophysics, Vrije Universiteit, Amsterdam, The Netherlands}
\address{$^c$Faculteit der  B\`{e}tawetenschappen, Vrije Universiteit, Amsterdam, The Netherlands}
\date{\today}

\begin{abstract}
We discuss the electroproduction of pseudoscalar ($0^{-+}$) or scalar ($0^{++}$) meson production off the scalar target.
The most general formulation of the differential cross section for the $0^{-+}$ or $0^{++}$ meson production process involves only 
one or two hadronic form factors, respectively, on a scalar target. The Rosenbluth-type separation of the differential cross section 
provides the explicit relation between the hadronic form factors and the different parts of the differential cross section in a completely
model-independent manner.
The absence of the beam spin asymmetry for the pseudoscalar meson production provides a benchmark for the experimental data analysis. The measurement of the beam spin asymmetry for the scalar meson production may 
also provide a unique opportunity not only to explore the imaginary part of the hadronic amplitude in the general formulation but also to examine the significance of the chiral-odd generalized parton distribution (GPD) contribution
in the leading-twist GPD formulation.
\end{abstract}

\maketitle

\section{Introduction}
\label{Sec1}
While the virtual Compton scattering process is 
coherent with the Bethe-Heitler process, 
the meson electroproduction process offers a unique experimental determination of the hadronic structures  
for the study of QCD and strong interactions. In particular, 
coherent electroproduction of  
pseudoscalar ($0^{-+}$) or scalar ($0^{++}$) mesons  off
a scalar target (e.g. the $^4$He nucleus) 
provides an excellent experimental terrain to discuss the fundamental nature of the hadron physics without
involving much complication from the spin degrees of freedom. 

We discuss in this work 
two benchmark examples ($0^{-+}$ vs. $0^{++}$) that provide a unique interface between the theoretical framework and the experimental measurements of physical observables.   

The paper is organized as follows. In Sec.~\ref{Sec2}, we summarize the formalism for the
electroproduction of pseudoscalar ($0^{-+}$) or scalar ($0^{++}$) meson off the scalar target.
In Sec.~\ref{Sec3}, we present the Rosenbluth-type separation of the differential cross section for the electroproduction 
of the $0^{-+}$ and $0^{++}$ mesons, from which the corresponding meson form factors can be directly extracted from the 
experimental data. In particular, we discuss the beam spin asymmetry (BSA) of the coherent meson 
($0^{-+}$ vs. $0^{++}$) electroproduction off the scalar target as well as the chiral-even vs. chiral-odd generalized parton 
distribution (GPD) contribution in the leading-twist GPD formulation. Summary and conclusion follows in Sec.~\ref{Sec4}.
In the Appendix, the evaluation of the scaling behaviors of chiral-even and chiral-odd amplitudes discussed in 
Sec.~\ref{Sec3} is briefly summarized.

\section{ General Formalism of Meson Electroproduction off the Scalar Target}
\label{Sec2}

\subsection{Cross Section and Invariant Amplitude Squared}
\begin{figure}
\includegraphics[height=6cm, width=7cm]{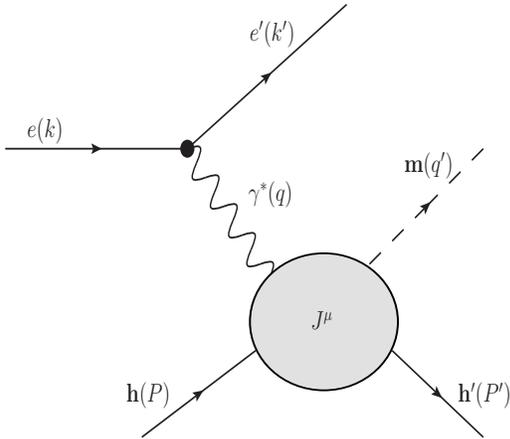}
\caption{\label{fig1} Momentum assignments in the meson electroproduction process with one-photon-exchange.}
\end{figure}

To establish the notation for the electroproduction of meson ${\bf m}$ off the scalar target ${\bf h}$, we write
\be\label{eq1}
e(k) + {\bf h}(P)  \to e'(k') + {\bf h}'(P') + {\bf m}(q'),
\ee
and the virtual photon momentum is defined to be $q = k- k'$, see Fig.~\ref{fig1}.

\begin{figure}
\includegraphics[height=6cm, width=9cm]{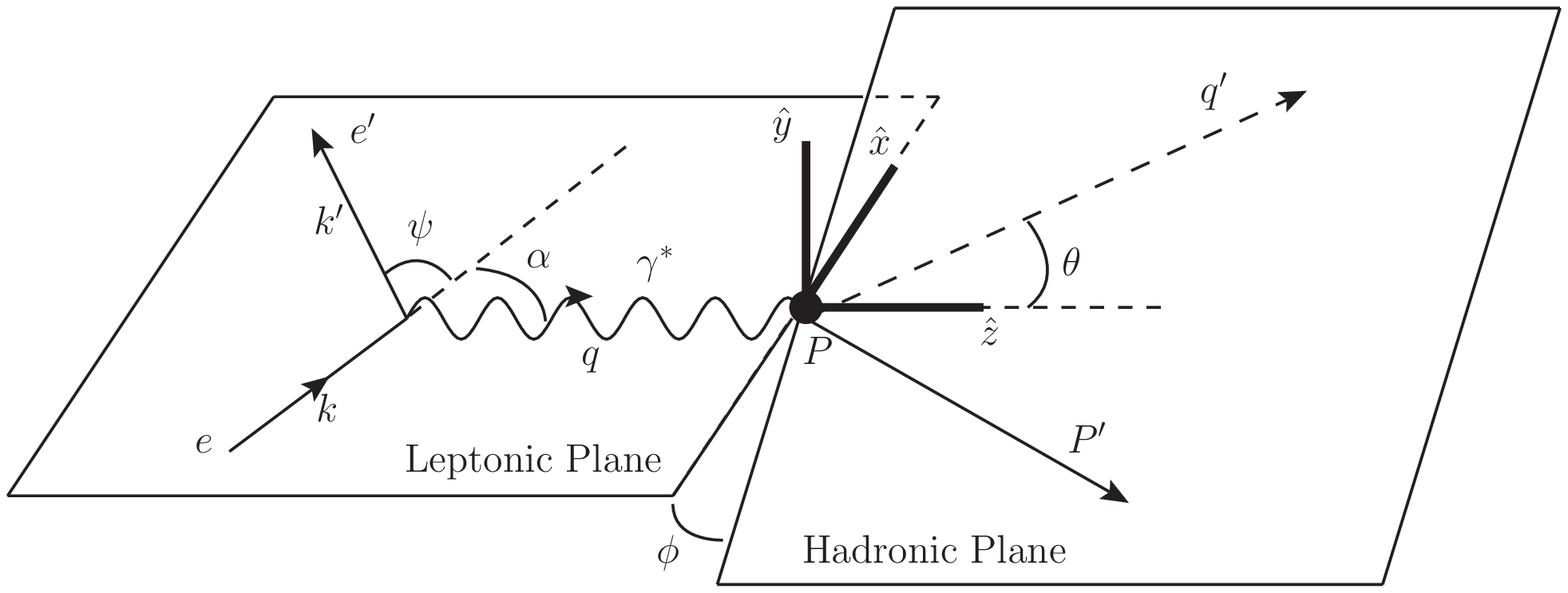}
\caption{\label{fig2} Target rest frame kinematics for the meson electroproduction.}
\end{figure}

In the target rest frame (TRF) presented in Ref.~\cite{WJC}, the 5-fold differential electroproduction cross section is given by
\bea\label{eq2}
d\sigma\equiv\frac{d^5\sigma}{ dy dx dt  d\phi_{k'} d\phi_{q'} } 
= \kappa \la |{\mathscr  M}|^2\ra,
\eea
where
\be\label{eq3}
\kappa \equiv \frac{1}{(2\pi)^5} \frac{y x}
{32 Q^2 \sqrt{1 + (\frac{2M x}{Q})^2}}.
\ee
Here, $y=P\cdot q/P\cdot k$, $t=(P-P')^2$ and $x=Q^2 /(2P\cdot q)=Q^2/(2M\nu)$ with $Q^2=-q^2$, the target mass $M$ and the virtual photon energy $\nu$ in TRF.
For the one-photon-exchange process, the transition amplitude ${\mathscr M}$ can be expressed as the invariant product of the leptonic current  $e L^\mu=e {\bar u}_{e'} (k',s') \gamma^\mu u_e (k,s)$  and the hadronic current $e J^\mu$ mediated by the photon propagator, i.e. ${\mathscr M} = e^2 L\cdot J/q^2$. As discussed in Ref.~\cite{WJC},  by using the reduced three momenta product obtained from the $q\cdot J=0$ relation, 
we get the following invariant amplitude squared
\bea\label{eq4}
\la |{\mathscr M}|^2 \ra &=&   \frac{e^4}{q^4} {\cal L}^{\mu\nu} {\cal H}_{\mu\nu}  \nonumber\\
&=&  \frac{e^4}{q^4} \biggl [
\frac{2 q^2}{\epsilon -1}\la |\tau_{fi}|\ra^2 
 + 2 i \lambda \epsilon^{\mu\nu \alpha\beta} 
 %\nonumber \\ 
 %&~~&\hspace{2cm} \otimes ~ 
 k_\alpha k'_\beta  J^{\dagger}_\mu J_\nu
\biggr ],\nonumber\\
\eea
where the hadronic tensor is given by
\be\label{eq4-2}
{\cal H}_{\mu\nu} = J^{\dagger}_\mu J_\nu,
\ee
and the leptonic tensor including the electron beam polarization $\lambda$ is given by
% \footnote{{\color{red}  
%If the spin sum is made for the initial electron, the factor 1/2 should be multiplied in the r.h.s. of Eq. (\ref{eq4-1}).}} 
%
\be\label{eq4-1}
{\cal L}^{\mu\nu} =q^2 \Lambda^{\mu\nu} + 2i \lambda \epsilon^{\mu\nu\alpha\beta} k_\alpha k'_\beta,
\ee
with $\Lambda^{\mu\nu}= g^{\mu\nu}+\frac{2}{q^2} (k^\mu k'^\nu + k'^\mu k^\nu)$.
Here, ${\cal L}^{\mu\nu}$ and ${\cal H}_{\mu\nu}$ are contracted to yield  Eq.~(\ref{eq4}) with
\bea\label{eq5}
\la |\tau_{fi}|\ra^2 &=& \frac{1}{2} ( |H_x|^2 + |H_y|^2) + \frac{\epsilon}{2}  ( |H_x|^2 - |H_y|^2) \nonumber \\
&+&\epsilon_L |H_z|^2 - \sqrt{\frac{1}{2} \epsilon_L (1+\epsilon)} 
%\nonumber \\
%&~~& \otimes ~~ 
(H^*_x H_z + H^*_z H_x),
\nonumber\\
\eea
where $\epsilon = \frac{\Lambda^{xx}-\Lambda^{yy}}{\Lambda^{xx}+\Lambda^{yy}}
=-\frac{2M^2 x^2 y^2+2Q^2(y-1)}{2M^2 x^2 y^2 + Q^2(y^2-2y +2)}$ and
$\epsilon_L = \frac{Q^2}{\nu^2} \epsilon$ and $H_i = J_i (i=x,y,z)$.
Typically in the laboratory, the kinematics of TRF depicted in Fig.\ref{fig2} is used. 
The angle $\psi$ in Fig.\ref{fig2} is related to $Q^2$ and $\nu$ as well as the beam energy $E$,
i.e. $Q^2 = -2 E(E-\nu)(1-{\rm cos}\,\psi)$. 
In terms of the angle $\psi$, the polarization parameter $\epsilon$ is given by
\be\label{eqPol}
\epsilon = \frac{1}{1+\frac{2 (\nu^2+Q^2)}{Q^2} {\rm tan}^2 \frac{\psi}{2}},
\ee
where one may note its consistency with Eq. (16) of Ref.~\cite{WJC} as well as 
${\rm tan}^2\frac{\psi}{2}=\frac{\epsilon_L(1-\epsilon)}{2\epsilon (\epsilon + \epsilon_L)}$
using $\epsilon_L/\epsilon = Q^2/\nu^2$. Also, neglecting the electron mass, one may note 
that the angle $\alpha$ between the beam (i.e. incident electron) direction and the virtual photon direction 
is related to $Q^2, \nu$ and $E$ as ${\rm cos}\,\alpha = \frac{Q^2 + 2E \nu}{2 E \sqrt{\nu^2+Q^2} }$.
The last terms in Eqs.~(\ref{eq4}) and (\ref{eq4-1}) for the case of a polarized electron beam
with $\lambda=\pm 1$ depending on the electron spin are related with the BSA.
Due to the absence of the interference with the Bethe-Heitler process, the BSA of the meson electroproduction 
is a direct measure of any asymmetry within the hadronic tensor, i.e., ${\cal H}_{\mu\nu} \neq {\cal H}_{\nu\mu}$.

\subsection{DNA Method for Hadronic Currents}
In parallel to the Levi-Civita symbol $\epsilon^{\mu\nu\alpha\beta}$, we have recently introduced in Ref.~\cite{JB2017}
the backbone of the Compton tensor defined by
\begin{equation}\label{eq6}
 d^{\mu\nu\alpha\beta} =
 g^{\mu\nu} g^{\alpha\beta} - g^{\mu\alpha} g^{\nu\beta},
\end{equation}
which may be used to construct pieces of ``DNA" for the virtual Compton scattering as well as the meson electroproduction by contracting with the three basis four vectors such as $q, {\bar P}=P + P'$ and  $\Delta = P - P' = q'-q$. The most general hadronic tensor structures obtained by our ``DNA" method
in virtual Compton scattering off the scalar target are in complete agreement with the previous results by Metz~\cite{Metz} and further comparisons with other methods~\cite{Eichmann} and results of general hadronic tensors for the nucleon target~\cite{NucleonTensorStructures} are underway.    

In the present work of the meson electroproduction off the scalar target, we note that the hadronic current for the pseudoscalar ($0^{-+}$) meson production is governed by a single 
hadronic form factor defined by 
\be\label{eq7}
J^\mu_{PS} = F_{PS} \epsilon^{\mu\nu\alpha\beta} q_{\nu} {\bar P}_\alpha \Delta_\beta,
\ee
while the hadronic current for the scalar ($0^{++}$) meson production involves two hadronic form factors
defined by
\be\label{eq8}
J^\mu_{S} = (S_q q_\alpha +
 S_{\bar P} {\bar P}_\alpha) d^{\mu\nu\alpha\beta} q_\beta \Delta_\nu,
\ee
where the hadronic form factors $F_{PS}$, $S_q$ and $S_{\bar P}$ are dependent 
on the Lorentz invariant variables $Q^2$,  $x$ and $t=\Delta^2$.
Defining the scalar hadronic form factors $F_1$ and $F_2$ for later convenience as 
%% end of change
\begin{eqnarray}\label{eq9}
F_1 = S_q - S_{\bar P}, &&
\nonumber \\
F_2 = S_{\bar P},&&
\label{eq.02n}
\end{eqnarray}
we get the hadronic current for the scalar ($0^{++}$) meson production as 
\bea\label{eq10}
J^\mu_{S} &=& F_1 (q^2 \Delta^\mu -  q\cdot\Delta\, q^\mu) \nonumber\\
&&+ F_2 [({\bar P}\cdot q + q^2) \Delta^\mu - q\cdot\Delta\,({\bar P}^\mu + q^\mu)],
\eea
which reduces to the usual electromagnetic current $J^\mu \propto (P+P')^\mu$ for the case of no meson production, i.e., $q' = 0$.
The electromagnetic current conservation is assured of course both for the electroproduction of pseudoscalar ($0^{-+}$) and 
scalar ($0^{++}$) mesons owing to $q_\mu J^\mu_{PS} = 0$ and $q_\mu J^\mu_{S} = 0$, respectively. 

%\vspace{0.5cm}
\section{Rosenbluth-type Separation of the Differential Cross Section and Beam Spin Asymmetry}
\label{Sec3}

\subsection{Pseudoscalar ($0^{-+}$) Meson Production Case}

For the pseudoscalar meson production case, we should note that the BSA term is zero because, owing to the fact that only a single hadronic form factor occurs, the hadronic tensor is symmetric: 
\bea\label{eq11}
{\cal H}_{\mu\nu} &=& 
|F_{PS}|^2 \epsilon_{\mu\alpha\beta\gamma} \epsilon_{\nu\alpha'\beta'\gamma'} 
q^\alpha {\bar P}^\beta \Delta^\gamma 
q^{\alpha'} {\bar P}^{\beta'} \Delta^{\gamma'} 
\nonumber \\
&=& {\cal H}_{\nu\mu},
\eea
and contracts with the antisymmetric leptonic tensor $2i \lambda \epsilon^{\mu\nu\alpha\beta} k_\alpha k'_\beta$ for
the BSA given by Eq.~(\ref{eq4}), i.e. 
\be\label{eq12}
\epsilon^{\mu\nu\alpha\beta} k_\alpha k'_\beta {\cal H}_{\mu\nu} = 0.
\ee
The situation here is very different from  $\pi^0$ electroproduction off a proton target in which several hadronic form factors are involved. The status of the data and phenomenology in the GPD approach of deeply virtual meson production (DVMP) on the nucleon has been reviewed in Ref.~\cite{DVMP}. The GPD formulation has been applied to the deeply virtual Compton scattering (DVCS) process off the pion~\cite{BMKS},
on spinless nuclear targets in the impulse approximation~\cite{Strikman} as well as off nuclei up to spin-1~\cite{Kirchner-Mueller}, and further refined 
for a spinless 
target~\cite{Belitsky-Mueller}. The coherent vs. incoherent DVCS processes off  spin-0 nuclei have also been discussed with respect to the nuclear medium modification of hadrons in terms of the GPD formulation~\cite{Liuti-Taneja-2005}.
In clear distinction from the recent BSA measurement of DVCS off $^4$He~\cite{Hattawy2017}, however,
the meson electroproduction process discussed here doesn't have any interference with the Bethe-Heitler process. 

As far as a single hadronic form factor governs the hadronic current, the BSA of the meson electroproduction should vanish in general
regardless of the complexity in the hadronic form factor. We thus note that the BSA of the coherent pseudoscalar (e.g. $\pi^0$) meson electroproduction off a scalar target (e.g. the $^4$He nucleus) vanishes due to the symmetry given by Eq.~(\ref{eq12}): i.e.,
\be\label{eq13}
\frac{d\sigma_{\lambda=+1}^{PS} - d\sigma_{\lambda=-1}^{PS}}{d\sigma_{\lambda=+1}^{PS} + d\sigma_{\lambda=-1}^{PS}} = 0.
\ee

Moreover, in the TRF kinematics~\cite{WJC} defining the azimuthal angle $\phi$ between the leptonic plane
and the hadronic plane taking the virtual photon direction as the ${\hat z}$-direction, the hadronic current for the pseudoscalar ($0^{-+}$) meson production given 
by Eq.~(\ref{eq7}) yields $H_z=0$ in Eq.~(\ref{eq5}). Regardless of the electron beam polarization $\lambda$,  
the differential cross section for the pseudoscalar meson (e.g. $\pi^0$) production is thus given by
\bea\label{eq14}
d\sigma^{PS}= d\sigma^{PS}_T + d\sigma^{PS}_{TT} \epsilon  \cos2\phi
= d\sigma^{PS}_T (1 - \epsilon  \cos2\phi),
\nonumber\\
\eea
where
\bea\label{eq14-1}
d\sigma^{PS}_T &=&-d\sigma^{PS}_{TT}
\nonumber\\
&=&\kappa  \frac{e^4 |F_{PS }(Q^2,t,x)|^2 \sin^2\theta}{4 M^2 x^4 (1-\epsilon)} 
 \left( 4 M^2 x^2+Q^2\right) \nonumber \\
 &&\times \left[ x^2 \left(t^2-4 m^2 M^2\right)+Q^4+2 Q^2 t x\right],
\eea
with the meson mass $m$ and the lab angle $\theta$ for the meson production in the hadronic plane. 
Here, we use the MAID notation~\cite{MAID} of the 5-fold differential cross section.
This provides the Rosenbluth-type separation 
of the differential cross section for the electroproduction of the pseudoscalar meson, 
allowing 
the pseudoscalar meson form factor $F_{PS}(Q^2,t,x)$ 
to be extracted directly from the 
experimental data of the differential cross section if available.

%\vspace{0.5cm}
\subsection{Scalar ($0^{++}$) Meson Production Case}

For the scalar meson production case, however, the BSA term doesn't vanish as there are two independent hadronic form factors 
$F_1(Q^2,t,x)$ and $F_2(Q^2,t,x)$ 
defined 
by Eq.~(\ref{eq10}), which are complex in general. 
The differential cross section for the scalar meson production is given by
\bea\label{eq15}
d\sigma^S_\lambda &=& d\sigma^S_T (1 + \epsilon  \cos(2\phi) ) + d\sigma^S_L \epsilon_L \nonumber \\
&& + ~ d\sigma^S_{LT}  \cos\phi \sqrt{\frac{1}{2}\epsilon_L (1+\epsilon)} 
+ \lambda \, d\sigma^S_{\rm BSA},
\eea
where $d\sigma^S_T  = d\sigma^S_{TT}$ and 
can be  written in terms of the form factors: 
\begin{gather}\label{eq16}
 \begin{bmatrix} 
 d\sigma^S_T \\ d\sigma^S_L \\ d\sigma^S_{LT} \\ d\sigma^S_{\rm BSA}
 \end{bmatrix}
 =
  \begin{bmatrix}
   T_1 & T_2 & T_3 & 0 \\
  L_1 &  L_2 & L_3 & 0 \\
  I_1  & I_2   & I_3 & 0 \\
  0 & 0 & 0& S_A
   \end{bmatrix}
   \begin{bmatrix} 
 |F_1|^2 \\ |F_2|^2 \\ F^+_{12} \\ F^-_{12}
 \end{bmatrix}
\end{gather}
with $F^\pm_{12} =F_1 F^*_2 \pm F_2 F^*_1$.
The matrix elements in Eq.~(\ref{eq16}) are obtained as follows:
\begin{widetext}
\bea\label{eq17}
T_1 &=& \frac{\kappa e^4\sin^2\theta Q^2}{4 M^2 x^2 (1-\epsilon )} 
 \left [x^2 \left(t^2-4 m^2 M^2\right)+Q^4+2 Q^2 t x\right],
\nonumber\\
T_2 &=& 
\frac{\kappa e^4 \sin^2\theta Q^2 (x-1)^2 }{4 M^2 x^4 (1-\epsilon )} 
\left[ x^2 \left(t^2-4 m^2 M^2\right)+Q^4+2 Q^2 t x\right],
\nonumber\\
T_3 &=& \sqrt{T_1 T_2},
\nonumber\\
L_1 &=& \frac{\kappa e^4 Q^4}{8 M^2 x^2 (1-\epsilon) \left( 4 M^2 x^2+Q^2\right)}
 \left[ m^2+Q^2+t (2 x-1)\right]^2,
\nonumber\\
L_2 &=&
\frac{\kappa e^4 \left[ m^2 \left( 4 M^2 x+Q^2\right)+Q^2 \left(4 M^2 x+2 t x - 3 t\right)-4 M^2 t x
+Q^4\right]^2}{8 M^2 x^2 (1-\epsilon) \left(4 M^2 x^2+Q^2\right) },
\nonumber\\
L_3 &=& \sqrt{L_1 L_2},
\nonumber\\
I_1 &=& 
\frac{\kappa e^4 I_c \tan\theta Q^2\left[ m^2+Q^2+t (2 x-1)\right] }
{2 M^2 x^2  (\epsilon -1) \left(4 M^2 x^2+Q^2\right)},
\nonumber\\
I_2 &=&
\frac{\kappa e^4 I_c  \tan\theta (x-1) }
{2  M^2 x^3  (\epsilon -1) \left(4 M^2 x^2+Q^2\right) }
\nonumber\\
&&\times
\left[m^2 \left(4 M^2 x+Q^2\right)
+Q^2 \left(4 M^2 x +2 t x-3 t\right)
-4 M^2 t x+Q^4
\right],
\nonumber\\
I_3&=&
\frac{\kappa e^4 I_c \tan\theta}{4 M^2 x^3 (\epsilon -1) \left(4 M^2 x^2+Q^2\right)}
 [ m^2 \left(4 M^2 x^2+Q^2 (2 x-1)\right)
 \nonumber\\
&& 
+Q^2 (4 M^2 x^2+4 t x^2-6 t x +t)
-4 M^2 t x^2+Q^4 (2 x-1) ],
\nonumber\\
S_A  &=&  - \kappa e^4 \frac{ \sin\theta  \sin\phi}{2 M x^2 y}
\left(m^2+Q^2-t\right) 
\nonumber\\
&\times&
 \sqrt{Q^2 (y-1) + M^2 x^2 y^2} \sqrt{x^2 \left(t^2-4 m^2 M^2\right)+Q^4+2 Q^2 t x},
\eea
%\end{widetext}
where
$I_c=2 M^2 x^2 \left(t-m^2\right)+Q^2 x \left(2 M^2 x+t\right)+Q^4$ and 
$\cos\theta = \frac{I_c}{Q\sqrt{(4 M^2 x^2 + Q^2)[x^2 (t^2 - 4m^2 M^2) + Q^4 + 2 Q^2 t x]}}$.

Thus, the BSA of the coherent scalar meson electroproduction off the scalar target is given by
\be\label{eq18}
\frac{d\sigma_{\lambda=+1}^{S} - d\sigma_{\lambda=-1}^{S}}{d\sigma_{\lambda=+1}^{S} + d\sigma_{\lambda=-1}^{S}} 
=\frac{d\sigma^S_{\rm BSA}}{d\sigma^S_T (1 + \epsilon  \cos(2\phi)) + d\sigma^S_L \epsilon_L
+ d\sigma^S_{LT}  \cos\phi \sqrt{\frac{1}{2}\epsilon_L (1+\epsilon)}},
\ee
\end{widetext}
which is proportional to $F_1 F^*_2 - F_2 F^*_1$. As $F_1 F^*_2 - F_2 F^*_1 \neq 0$ in general, 
the BSA of the scalar meson (e.g. $f_0(980)$) electroproduction is not expected to vanish.
For the kinematic region where at least one of $F_1$ or $F_2$ develops an imaginary part,
the BSA shouldn't vanish. The nonvanishing BSA measured in DVCS off $^4$He~\cite{Hattawy2017} 
indicates that the imaginary part of the hadronic amplitude is accessible in the current experimental regime. 
Therefore, it will be very interesting to compare the experimental data on the BSAs between the $\pi^0$ electroproduction
and the $f_0(980)$ electroproduction off the $^4$He nucleus. We note that Eqs.~(\ref{eq15}) - (\ref{eq17}) 
provide the Rosenbluth-type separation of the differential cross section for the electroproduction of the scalar meson, 
allowing
the scalar meson form factors $F_1 (Q^2,t,x)$ and $F_2 (Q^2,t,x)$ to be directly extracted from the 
experimental data. In principle, the experimental data can reveal both the real part and the imaginary part of 
$F_1 (Q^2,t,x)$ and $F_2 (Q^2,t,x)$ through Eqs.~(\ref{eq15}) - (\ref{eq17}) and the consistency with 
the BSA given by Eq.~(\ref{eq18}) can be checked for the kinematic region where any of these form factors 
is found to develop an imaginary part.

%\vspace{0.5cm}
\subsection{Comparison with the GPD formulation}

The leading-twist GPD formulation~\cite{XJI1997,AVR1996,Mueller94}
provides detailed information of the individual contribution from each and every constituent of the target. 
The most well-known example of GPD formulation may be found in DVCS for the proton target which has 
four twist-2 GPDs ($H, E, {\tilde H}, {\tilde E}$).
Regardless of DVCS or DVMP, the GPD formalism  relies on the ``handbag dominance" 
representing the factorization of 
the hard and soft parts
in the respective scattering amplitudes. 
It is well known that the integrals of the leading-twist GPDs in the $s$- and $u$-channel handbag amplitudes of 
both DVCS and DVMP processes carry the factorized denominator factors such as $1/(x-\zeta)$ and $1/x$, respectively. 
Here, $\zeta = \Delta^+ / P^+$ is the skewness variable in the GPD formulation~\cite{XJI1997, AVR1996}
and $x = k^+/P^+$ is the light-front longitudinal momentum fraction of the particle with the four-momentum $k^\mu$ struck by the probing virtual photon with respect to $P^+$. The kinematic region for the handbag dominance
is typically provided by $\left|t\right| \ll Q^2$~\cite{JBreview,BJPRD(R)}. 

Our findings from the general formulation with two independent hadronic form factors for the electroproduction of the scalar ($0^{++}$) meson may be compared with the GPD formulation discussed in the reviews\cite{B&R, Diehl} 
which provided the number of leading-twist GPDs for the same process. In particular, one should note that not only 
the chiral-even operator $\gamma^+$ but also the chiral-odd operator $\sigma^{+\perp}_\mu$ can be effective for 
spin-zero hadrons, providing  the contribution from the two twist-2 GPDs, i.e., the chiral-even GPD ($H$) and 
the chiral-odd GPD ($H_T$), respectively, to the DVMP process of the scalar ($0^{++}$) meson production.
As pointed out in Ref.~\cite{B&R}, the GPDs defined by the aligned parton-helicity operators are allowed 
due to nonzero orbital angular momentum between the initial and final state hadrons.  
One may check explicitly the helicity flip vs. non-flip amplitudes
in the quark level including not only the identity coupling to the quark-scalar ($0^{++}$) meson vertex 
which singles out the chiral-odd GPD ($H_T$) but also the derivative coupling with $\gamma_\mu$ to 
the quark-scalar ($0^{++}$) meson vertex which provides the chiral-even GPD ($H$) contribution.
As the chirality and the helicity coincides in the massless limit, it's rather straightforward to identify 
the chiral-even vs. chiral-odd contribution from the helicity flip vs. non-flip amplitudes, respectively.
From the evaluation of helicity flip vs. non-flip amplitudes 
as discussed in the Appendix, 
one may realize that the derivative 
coupling with $\gamma_\mu$ can bring $\sim \sqrt{Q^2}$ over the non-derivative identity coupling. 
While this might naively suggest the $\sqrt{-t}/Q$ suppression of
the chiral-odd contribution with respect to the chiral-even contribution, one should note that very little is 
known on the scalar ($0^{++}$) meson wave function in the quark-scalar ($0^{++}$) meson vertex. 
Overcoming the $\sqrt{-t}/Q$ factor, if the chiral-odd GPD ($H_T$) contributes as significantly as 
the chiral-even GPD ($H$), then the GPD formulation would provide the nonvanishing BSA 
in DVMP of scalar ($0^{++}$) meson production off the scalar target as we have discussed 
with the two independent hadronic form factors in Eqs.(\ref{eq15})-(\ref{eq18}).
By the same token, the experimental observation of the nonvanishing BSA of a scalar meson (e.g. $f_0(980)$) electroproduction off a scalar target (e.g. the $^4$He nucleus) would reveal a remarkable 
chiral-odd GPD ($H_T$) contribution in the leading-twist GPD formulation. Unless the chiral-odd GPD ($H_T$)
contributes as significantly as the chiral-even GPD ($H$), a single GPD contribution alone would provide 
a zero BSA, $d\sigma^S_{BSA} = 0$. 

It is also important to note that the BSA requires a non-zero $t$ as it is defined in terms of 
the azimuthal angle $\phi$. As it has been shown in Ref.~\cite{Hattawy2017}, 
the measurement of the BSA in the kinematic region $\left|t\right| \ll Q^2$ can still be analyzed without involving any 
higher-twist GPDs. While the BSA measurement presented in Ref.~\cite{Hattawy2017} was  
restricted to the kinematic region $\left|t\right| \ll Q^2$, the experimental data were analyzed with the single leading-twist 
GPD, $H_A$, only\footnote{As discussed in Ref.~\cite{JBreview}, the number of Compton form factors in virtual Compton scattering off a scalar target is three. Our work including both Bethe-Heitler process and virtual Compton scattering process off a scalar target is underway.}.

%\vspace{0.5cm}
\section{Summary and Conclusion}
\label{Sec4}

In summary, either a single hadronic form factor or a single leading-twist GPD would result in the symmetric hadronic tensor ${\cal H}_{\mu\nu} = {\cal H}_{\nu\mu}$ as we have discussed in the case of  pseudoscalar meson electroproduction. 
This would then yield a vanishing BSA as the symmetric hadronic tensor ${\cal H}_{\mu\nu}$ contracts with the antisymmetric leptonic tensor $2i \lambda\epsilon^{\mu\nu\alpha\beta} k_\alpha k'_\beta$. 
The absence of the beam spin asymmetry for the pseudoscalar meson production provides a benchmark for the experimental data analysis. 

Not only the pseudoscalar meson production but also the scalar meson production provides benchmark results for the interface between the theoretical framework and the experimental measurements of physical observables.   
The coherent experimental measurement to judge whether the BSA of a scalar meson (e.g. $f_0(980)$) electroproduction off a scalar target (e.g. the $^4$He nucleus) vanishes or not would provide a unique opportunity to 
explore the imaginary part of the hadronic amplitude accessible in the general formulation
with the two independent hadronic form factors, $F_1$ and $F_2$. It would also examine the significance of whether
the chiral-odd GPD ($H_T$) contribution is on par with the chiral-even GPD ($H$) contribution in the leading-twist 
GPD formulation. 

In this respect, both pseudoscalar and scalar meson electroproduction measurements off a scalar target are highly desired to pin down the viable roadmap on the analyses of precision experimental data, e.g. from the JLab 12 GeV upgrade. An exactly solvable hadronic model calculation is currently underway to explore the kinematic regions where 
the hadronic form factors develop imaginary parts, and to explicitly demonstrate the extraction of the hadronic form factors
from our general formulation of the hadronic currents. The recent report on the experimental studies of DVMP and 
transversity GPDs~\cite{Kubarovsky} attracts our attention to the nucleon target as well.    

\section*{Acknowledgments}
We thank Prof. Kyungseon Joo for useful discussions on meson electroproduction processes.    
This work was supported by the U.S. Department of Energy (Grant No. DE-FG02-03ER41260). 
CRJ thanks the Jefferson Lab Theory Center for the hospitality during his visit in the midst of completing this work.
HMC's work was in part supported by the National Research Foundation of Korea (NRF) (Grant No. NRF-2017R1D1A1B03033129).

%\newpage
%\begin{widetext}
\appendix*
\section{SCALING BEHAVIORS OF CHIRAL-EVEN AND CHIRAL-ODD AMPLITUDES}
We use here the following kinematics~\cite{JBreview} in electroproduction of meson off the massless quark:
\begin{eqnarray}
k^\mu &=& (x p^+,0,0,0), \nonumber \\
k'^\mu &=&\biggl ((x-\zeta)p^+,-\Delta_\perp,0,-\frac{\Delta_\perp^2}{2(x-\zeta)p^+} \biggr),\nonumber \\
q^\mu &=& \biggl((-\zeta + \alpha (\mu_s + \mu_t))p^+,0,0,\frac{Q^2}{2 p^+}(\frac{1}{\alpha}+\frac{\mu_t}{x-\zeta}) \biggr),\nonumber \\
q'^\mu &=& \biggl(\alpha(\mu_s+\mu_t)p^+,\Delta_\perp,0,\frac{Q^2}{2\alpha p^+}\biggr),
\end{eqnarray}
where $k(k')$ and $q(q')$ are the momenta of the incoming(outgoing) quarks and photons, respectively and
where $\Delta^\mu = q'^\mu - q^\mu = k^\mu - k'^\mu$, $\mu_s =m^2/Q^2$, $\mu_t =\Delta_\perp^2/Q^2$ and
\begin{eqnarray}
\alpha &=& \frac{2\zeta (x-\zeta)}{((1+\mu_s+\mu_t)(x-\zeta)-\mu_t\zeta-\sqrt{D}}, 
\end{eqnarray}
with
%\begin{equation}
$D = 4\mu_t(\mu_s+\mu_t)\zeta(x-\zeta) +((1+\mu_s+\mu_t)(x-\zeta)-\mu_t\zeta)^2$.
We note that $\alpha \approx \zeta$ as $\mu_s \to 0$ and $\mu_t \to 0$.

The corresponding Mandelstam variables $s=(k+q)^2$, $t=(k-k')^2$, and $u=(k-q')^2$
are given by
%\begin{widetext}
\begin{eqnarray}
s &=&  \frac{Q^2}{2\zeta(x-\zeta)} \biggl((1+\mu_s+\mu_t)x^2-(3+\mu_s)x\zeta \nonumber\\
            &&\hspace{2cm} +2\zeta^2+x\sqrt{D} \biggr), \nonumber \\
t &=& \Delta^2 = -\frac{x}{x-\zeta}\Delta_\perp^2, \nonumber \\
u &=&\frac{-Q^2}{2\zeta(x-\zeta)} \biggl((1+\mu_s+\mu_t)x^2  - (1+3\mu_s+2\mu_t)x\zeta \nonumber\\
           &&\hspace{2cm} + 2\mu_s \zeta^2 + \sqrt{D} \biggr),
%s+t+u &=& -Q^2+m_s^2
\end{eqnarray}
and $s+t+u = -Q^2+m^2$.

%\end{widetext}
%\newpage
The hadronic amplitudes of the $S$ and $U$ channels in the quark level are respectively given by
\begin{eqnarray}
&&J_{h,{h'},\lambda}^S={\bar u}_{h'}(k-\Delta) \Gamma ({\not\!k}+{\not\!q}+m_q){\not\!\epsilon}_\lambda(q) u_h(k), \nonumber \\
&&J_{h,{h'},\lambda}^U={\bar u}_{h'}(k-\Delta) {\not\!\epsilon}_\lambda(q)\, ({\not\!k}-{\not\!{q'}}+m_q) \Gamma \, u_h(k), 
\nonumber\\
\end{eqnarray}
where the scalar meson vertex is generally taken as $\Gamma = E_S + F_S{\not\!{q'}} + G_S{\not\!k} + H_S \,\sigma^{\mu\nu}{q'_\mu}{k_\nu}$
with the external momentum $q'$ and the internal momentum $k$,
and the quark mass $m_q$ is taken to be zero after the calculation. \\
In the limit $m_q \rightarrow 0$, each hadronic amplitude can be expanded in the orders of $\frac{\Delta_{^\perp}}{Q}$ and the results up to 
the second order $\left(\frac{\Delta_{^\perp}}{Q}\right)^2$ 
are summarized as follows:
\begin{eqnarray}
J_{\uparrow,\uparrow,+1}^S&=&0, \nonumber \\
J_{\uparrow,\uparrow,-1}^S&=&-\sqrt{2} F_S \, Q^3 \sqrt{\frac{x}{x-\zeta}} \frac{\Delta_\perp}{Q},\nonumber \\
J_{\uparrow,\uparrow,0}^S&=&-i F_S\, Q^3 \sqrt{\frac{x}{x-\zeta}} \frac{\Delta_\perp^2}{Q^2+m^2}, \nonumber \\
J_{\uparrow,\downarrow,+1}^S&=&0, \nonumber \\
J_{\uparrow,\downarrow,-1}^S&=&-\sqrt{2} E_S\, Q^2 \sqrt{\frac{x}{x-\zeta}} \frac{\Delta_\perp^2}{Q^2+m^2}, \nonumber \\
J_{\uparrow,\downarrow,0}^S&=&-i E_S\, Q^2 \sqrt{\frac{x}{x-\zeta}} \frac{\Delta_\perp}{Q},
\end{eqnarray}
for the $S$ channel, and
\begin{eqnarray}
J_{\uparrow,\uparrow,+1}^U&=&0, \nonumber \\
J_{\uparrow,\uparrow,-1}^U&=&-\sqrt{2} F_S\, Q^3 \sqrt{\frac{x}{x-\zeta}} \left (\frac{m^2}{Q^2}\right) \frac{\Delta_\perp}{Q}, \nonumber \\
J_{\uparrow,\uparrow,0}^U&=&-i F_S\, Q^3 \sqrt{\frac{x}{x-\zeta}} \left (\frac{m^2}{Q^2} \right) \frac{\Delta_\perp^2}{Q^2+m^2}, \nonumber \\
J_{\uparrow,\downarrow,+1}^U&=&-\sqrt{2} E_S\, Q^2 \sqrt{\frac{x}{x-\zeta}}  \frac{\Delta_\perp^2}{Q^2}, \nonumber \\
J_{\uparrow,\downarrow,-1}^U&=&\sqrt{2} E_S\, Q^2 \sqrt{\frac{x}{x-\zeta}} \frac{\Delta_\perp^2}{Q^2+m^2}, \nonumber \\
J_{\uparrow,\downarrow,0}^U&=&i E_S\, Q^2 \sqrt{\frac{x}{x-\zeta}} \left (\frac{m^2}{Q^2}\right) \frac{\Delta_\perp}{Q}.
\end{eqnarray}
for the $U$ channel. We note here that $G_S$ and $H_S$ coming with the internal momentum $k$ do not appear in our leading order calculation.
These results show that the helicity flip and non-flip amplitudes contribute with the $E_S$ and $F_S$ terms 
in the scalar vertex $\Gamma$, respectively, where the $E_S$ term carries the identity operator while the
$F_S$ term carries the $\not\!{q'}$ operator which is dominated by the $\gamma^+$ operator with a $Q^2$ factor.  
In the large $Q$ limit of DVMP, the helicity non-flip (chiral even) amplitude dominates over the helicity flip (chiral odd) 
amplitude by one higher order of $Q$ as the leading order contributions of helicity flip and non-flip
amplitudes are given by $J_{\uparrow,\downarrow,0}^S = -i E_S \, Q^2 \sqrt{\frac{x}{x-\zeta}} \frac{\Delta_{^\perp}}{Q} \sim Q \Delta_\perp$
and $J_{\uparrow,\uparrow,-1}^S = -\sqrt{2} F_S \, Q^3 \sqrt{\frac{x}{x-\zeta}} \frac{\Delta_{^\perp}}{Q} \sim Q^2 \Delta_\perp$, respectively.
Consequently, in the GPD formulation, the chiral-odd GPD contribution appears to be suppressed by an order of $1/Q$ with respect to
the chiral-even GPD contribution, unless $E_S/F_S \sim Q$. 
%\end{widetext}


\begin{thebibliography}{18}
\bibitem{WJC} R. A. Williams, C.-R. Ji and S. R. Cotanch, \Journal{\PRC}{46}{1617}{1990}.

\bibitem{JB2017} C.-R. Ji and B.L.G. Bakker, PoS QCDEV2017, 038 (2017); B.L.G. Bakker and C.-R. Ji, 
Few Body Syst. {\bf 58}, 8 (2017).

\bibitem{Metz}
A. Metz, Virtuelle Comptonstreuung und die Polarisierbarkeiten des Nukleons (in German), PhD thesis, Universit\"at Mainz, 1997.

\bibitem{Eichmann} 
G. Eichmann, C. S. Fischer and W. Heupel, Phys.\ Rev.\ D \textbf{92}, 056006 (2015); 
G. Eichmann, ``Baryon spectroscopy and structure with functional methods", presentation in Light Cone 2018 at JLab, May 14-18, 2018; 
G. Eichmann and G. Ramalho, arXiv:1806.04579[hep-ph].

\bibitem{NucleonTensorStructures} D. Drechsel, G. Kn\"ochlein, A. Yu. Korchin, A. Metz and S. Scherer, Phys. Rev. C {\bf 57}, 941 (1998); 
R.Tarrach, Nuovo Cim A {\bf 28}, 409 (1975).

\bibitem{DVMP} L. Favart, M. Guidal, T. Horn and P. Kroll, Eur. Phys. J. A {\bf 52}, 158 (2016).

\bibitem{BMKS}
A.V. Belitsky, D. M\"uller, A. Kirchner and A. Sch\"afer, Phys. Rev. D {\bf 64}, 116002 (2001).

\bibitem{Strikman} V.~Guzey and M. Strikman,
Phys.\ Rev.\ C \textbf{68}, 015204 (2003).

\bibitem{Kirchner-Mueller}
A. Kirchner and D. M\"uller, Eur. Phys. J. C {\bf 32}, 347 (2003).

\bibitem{Belitsky-Mueller}
A.V. Belitsky and D. M\"uller, Phys. Rev. D {\bf 79}, 0141017 (2009). 

\bibitem{Liuti-Taneja-2005} S. Liuti and S. K. Taneja, \Journal{\PRC}{72}{034902}{2005}.

\bibitem{Hattawy2017} M. Hattawy et al. (CLAS Collaboration), Phys. Rev. Lett. {\bf 119}, 202004 (2017).

\bibitem{MAID} D. Drechsel and L. Tiator, J. Phys. G \textbf{18}, 449 (1992).

\bibitem{XJI1997}
X.~D.~Ji,
%``Gauge invariant decomposition of nucleon spin,''
Phys.\ Rev.\ Lett.\  \textbf{ 78}, 610 (1997);
Phys.\ Rev.\ D \textbf{ 55}, 7114 (1997).

\bibitem{AVR1996}
A.~V.~Radyushkin,
%``Scaling Limit of Deeply Virtual Compton Scattering,''
Phys.\ Lett.\ B \textbf{ 380}, 417 (1996);
A.V. Radyushkin, Phys. Rev. D {\bf 56}, 5524 (1997).

\bibitem{Mueller94}
D. M\"uller, D. Robaschik, B. Geyer, F. M. Dittes and J. Horejsi, Fortsch. Phys. \textbf{42}, 101 (1994).

\bibitem{JBreview} C.-R. Ji and B.L.G. Bakker,
Int. J. Mod. Phys. E {\bf 22}, 1330002 (2013). 


\bibitem{BJPRD(R)}
B.L.G. Bakker and C.-R. Ji,
Phys.\ Rev.\ D \textbf{83}, 091502(R) (2011).

\bibitem{B&R} A. V. Belitsky and A. V. Radyushkin, Phys. Rept. \textbf{418}, 1 (2005).

\bibitem{Diehl} M. Diehl, Phys. Rept. \textbf{388}, 41 (2003).

\bibitem{Kubarovsky} V. Kubarovsky (CLAS Collaboration), arXiv:1902.02643v1 [hep-ph].


\end{thebibliography}
\end{document}